\documentclass[    ,final      ]{aipproc}
\layoutstyle{6x9}
\usepackage{amsfonts,amsmath,amssymb,mathrsfs,graphicx}
\begin{document}

\title{Collapse and Relativity}
%The Ghirardi--Rimini--Weber Theory and Relativity (and Quantum
%   Field Theory)}

\author{Roderich Tumulka}{address ={Mathematisches Institut,
    Eberhard-Karls-Unversit\"at, Auf der Morgenstelle 10, 72076
    T\"ubingen, Germany.\\e-mail:
    tumulka@everest.mathematik.uni-tuebingen.de}
}

\begin{abstract}
Ever since we have been in the possession of quantum theories without observers, such as Bohmian mechanics or the Ghirardi--Rimini--Weber (GRW) theory of spontaneous wave function collapse, a major challenge in the foundations of quantum mechanics is to devise a relativistic quantum theory without observers. One of the difficulties is to reconcile nonlocality with relativity. I report about recent progress in this direction based on the GRW model: A relativistic version of the model has been devised for the case of $N$ noninteracting (but entangled) particles. A key ingredient was to focus not on the evolution of the wave function but rather on the evolution of the matter in space-time as determined by the wave function.
\end{abstract}

\maketitle

%\addtolength{\textwidth}{2.0cm}
%\addtolength{\hoffset}{-1.0cm}
%\addtolength{\textheight}{3.0cm}
%\addtolength{\voffset}{-1.5cm}

\newcommand{\RRR}{\mathbb{R}}
\newcommand{\CCC}{\mathbb{C}}
\newcommand{\PPP}{\mathbb{P}}
\newcommand{\ZZZ}{\mathbb{Z}}
\newcommand{\Hilbert}{\mathscr{H}}
\newcommand{\D}{\mathrm{d}} % differential d
\newcommand{\E}{\mathrm{e}} % exponential e
\newcommand{\I}{\mathrm{i}} % imaginary i

\section{Introduction}

I  describe some models I designed recently that generalize the Ghirardi--Rimini--Weber (GRW) model \cite{grw,Belljumps,BG03} of spontaneous wave function collapse. One of these \cite{Tum06a} is a relativistic version of the GRW model, the other \cite{Tum06b}, which I will describe first, an extension to quantum field theory (QFT). It would be nice to combine the two models into a single one suitable for relativistic quantum field theory; I have not started this task yet, but I hope I will be able to do this soon. In this article, I will begin by elucidating the flash ontology for the GRW model \cite{Belljumps}, which is crucial for understanding the two new models.

\section{The Flash Ontology}

\begin{figure}[ht]
\includegraphics[width=.4 \textwidth]{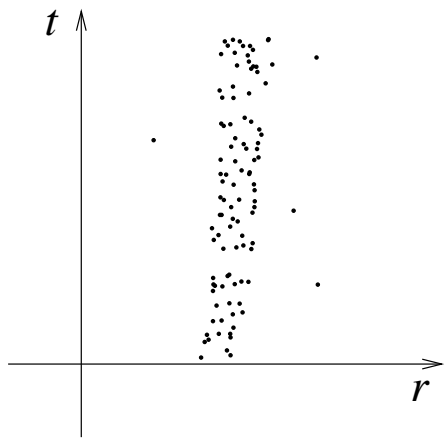}
\caption{A typical pattern of flashes in space-time, and thus a possible   
world according to the flash version of the GRW theory}
\label{flashes}
\end{figure}

\emph{What are tables and chairs made of?} To this question, different theories, even different versions of the GRW model, may propose different answers. Classical physics would say particles, described mathematically by their world lines. The objects that the answer describes, whatever it may be, were termed the ``primitive ontology'' of the theory by D\"urr, Goldstein, Zangh\`\i, and Allori \cite{DGZ04,All04,AGTZ} and the ``local beables'' by Bell \cite{Bell75}. To give you concrete examples to think of, let me introduce two possible answers that have been proposed for the GRW model:
\begin{itemize}
\item[A)] Not the wave function, but a continuous distribution of matter with density
\begin{equation}\label{mdef}
  m(r,t) = \sum_{i=1}^N \int \D^3 r_1 \cdots \widehat{\D^3r_i} \cdots \D^3r_N \,
  \Bigl| \psi(r_1, \ldots, \widehat{r_i},r, \ldots, r_N,t) \Bigr|^2 \,,
\end{equation}
where the hat denotes omission. This is (a simplified version of) a proposal due to Benatti, Ghirardi, and Grassi \cite{Ghi}.

\item[B)] Not the wave function, but discrete ``flashes'', i.e., elementary events represented by space-time points. This is a proposal due to Bell \cite{Belljumps,Bellexact,kent}. In such a world, matter consists of lots of dots in space-time (see Fig.~\ref{flashes}), and ``a piece of matter then is a galaxy of such events'' (Bell \cite{Belljumps}). Such a space-time point $X=(R,T)$ is defined in the GRW model as the center of a wave function collapse.
\end{itemize}

I will use the flash ontology in the following. It is misleading to say that the flashes \emph{are} the collapse centers, though the collapse centers determine where in space-time the flashes are. It is better to think of the flashes as what the theory is about; when simulating the GRW model with the flash ontology on a computer, the set of flashes is the \emph{output} of the algorithm. That is also why I prefer the word ``flash'' to ``hitting,'' which would suggest instead that the wave function be the output of the theory.
The right way to think of the primitive ontology is as describing the motion of matter in space-time, and thus as analogous to the particle trajectories in classical (or Bohmian) mechanics. A more thorough discussion of the notion of primitive ontology can be found in \cite{AGTZ}.

The flash ontology is an unusual choice of ontology; a more normal choice would be particle world lines, or fields in space-time. The motivation for this choice lies in the fact that the GRW model with flashes can be made Lorentz invariant (by suitable corrections in the equations), whereas the GRW model with the matter density $m(r,t)$ cannot in any known way. The GRW model without primitive ontology, where only the wave function is regarded as existing, suffers, I think, from serious problems \cite{AGTZ}. The basic problem is that a theory that never talks about matter would not form an adequate description of our world. Strictly speaking, in a world governed by such a theory there exists no matter, or else the link between matter and the mathematical variables of the theory is left vague. A related problem is that it remains to be clarified in what way the wave functions $\psi_\Sigma$ and $\psi_{\Sigma'}$ associated with different spacelike 3-surfaces $\Sigma$ and $\Sigma'$ have to be compatible to be regarded as describing a consistent reality. In case $\Sigma$ and $\Sigma'$ have a large portion in common, one would presumably require that $\psi_\Sigma$ and $\psi_{\Sigma'}$ describe the same reality on $\Sigma \cap \Sigma'$. But the heart of this problem is that one has not defined clearly what that reality is and how the wave function influences that reality, and to provide such a definition is exactly the purpose of the concept of primitive ontology.

When we regard GRW as a theory about flashes then the flashes form a random set, a \emph{point process} (as mathematicians would say), in space-time. The joint distribution of the flashes is then determined by the (initial) wave function. Indeed, for this distribution one can write an explicit formula, which I will do in the next section, albeit in much greater generality including quantum field theories.

\section{GRW and QFT}

It will be useful to write the abstract mathematical structure of the GRW model in the following way \cite{Tum06b}. The mathematical ingredients are:
\begin{itemize}
\item The initial wave function (at time $t_0$), a vector $\psi$ in a Hilbert space $\Hilbert$ with $\|\psi\|=1$.

\item For writing the flash rate (the probability per time with which a flash occurs) at a location $r \in \RRR^3$ in the form
\begin{equation}
  \langle{\psi}|\Lambda(r)|\psi\rangle \,,
\end{equation}
we need a positive self-adjoint operator $\Lambda(r)$ for every $r\in\RRR^3$. 

For the case of nonrelativistic quantum mechanics of $N$ distinguishable particles, it is useful to introduce $N$ types of flashes (corresponding to the $N$ variables $r_i \in \RRR^3$ of the wave function), and thus for every type $i\in \{1,\ldots, N\}$ the flash rate operators $\Lambda_i(r)$. To obtain the original GRW model, choose the flash rate operators to be multiplication operators by (3-dimensional) Gaussians,
\begin{equation}\label{GRWGauss}
  \Lambda_i(r) \, \psi(r_1,\ldots,r_N) = \frac{1}{\tau} \, \frac{1}{(2\pi \sigma^2)^{3/2}}\,
  \E^{-(r-r_i)^2/2\sigma^2} \, \psi(r_1, \ldots, r_N)\,,
\end{equation}
where the model constants $\sigma$ and $\tau$ are the width of the localization (say, $\sigma = 10^{-7}$ meters), and the average time between two collapses of the same type $i$ (say, $\tau = 10^8$ years).

\item The Hamiltonian $H$ defines the way $\psi$ would evolve if there were no collapses.
\end{itemize}

Together, $\Hilbert,H$, and $\Lambda$ define a flash theory by the postulate that the joint distribution of the first $n$ flashes is given by
\begin{equation}\label{ndistr}
  \PPP\bigl( X_1 \in \D^4x_1, \ldots, X_n \in \D^4x_n\bigr) =
  \|K_n \psi\|^2 \, \D^4x_1 \cdots \D^4x_n \,,
\end{equation}
where $\D^4x_k = \D^3r_k \, \D t_k$ is the 4-volume, $x_k=(r_k,t_k)$, and
\begin{equation}\label{Kdef}
  K_n = \Lambda(r_n)^{1/2} \, W_{t_n-t_{n-1}} \cdots \Lambda(r_1)^{1/2} \, W_{t_1-t_0}
\end{equation}
with
\begin{equation}\label{Wdef}
  W_t = \begin{cases}
  \exp\bigl( -\I Ht/\hbar - \tfrac{1}{2} \int \D^3r \, \Lambda(r) t \bigr) 
  & \text{if $t\geq 0$}\\
  0&\text{if $t<0$.}
  \end{cases}
\end{equation}
It is an easy exercise to check formally that with these definitions, \eqref{ndistr} consistently defines a probability distribution on the suitable history space (a history in the flash ontology is a pattern of flashes, and history space thus is the space of all discrete subsets of space-time). The original GRW model is a special case of this scheme, for which $\int \D^3r \, \Lambda(r)$ is, by \eqref{GRWGauss}, a multiple of the identity operator in $\Hilbert$, so that $W_t$ is the unitary Schr\"odinger evolution up to a scalar factor $\exp(-t/2\tau)$ that governs the (exponential) distribution of the random waiting time until the next flash. For the case of several types of flashes there exist obvious analogs to equations \eqref{ndistr}--\eqref{Wdef}, see \cite{Tum06b}. I also mention that equations \eqref{ndistr}--\eqref{Wdef} fit into a general mathematical scheme that Blanchard and Jadczyk \cite{BJ95} have developed for a different purpose, namely for models of coupled classical and quantum systems; see Jadczyk's contribution to this volume for details. The subsequent work \cite{BJ96,Ru02} on relativistic extensions of their approach, in contrast, seems not related to any relativistic theory of flashes.

For $N$ identical particles, one should form the operators
\begin{equation}\label{id}
  \Lambda^{(N)}(r) = \sum_{i=1}^N \Lambda_i(r)\,,
\end{equation}
restricted to the subspace of anti-symmetric (or, symmetric) wave functions for fermions (respectively, bosons). We thus obtain, in a remarkably simple way, a variant of the GRW model appropriate for identical particles, despite the suggestion sometimes conveyed in the literature (see, e.g., pages 312 and 382 in \cite{BG03}) that for dealing with identical particles one has to employ a ``CSL'' model (continuous spontaneous localization \cite{GPR90}), in which the state vector follows a diffusion process in Hilbert space and thus ``collapses continuously'' rather than at discrete times. In fact, the modification of the GRW collapse process corresponding to the choice \eqref{id} of flash rate operators was already proposed in 1995 by Dove and Squires \cite{DS95}, but has not received the attention it deserves.

For a variable number of particles, one should form
\begin{equation}
  \Lambda (r) = \bigoplus_{N=0}^\infty \Lambda^{(N)}(r)
\end{equation}
on Fock space. In a sense, quantum field theory is nothing but the quantum mechanics of a variable number of identical particles. This allows us to define a flash process for more or less every QFT that is regularized (i.e., possesses a well-defined Hamiltonian). Equivalently, we can write $\Lambda(r)$ as smeared-out ``particle number density'' operators $N(r)$,
\begin{equation}\label{LambdaN}
  \Lambda(r) =  \frac{1}{\tau} \int \D^3r'\, \frac{1}{(2\pi\sigma^2)^{3/2}} \, \E^{-(r-r')^2/2\sigma^2} \,
  N(r')\,.
\end{equation}
(In nonrelativistic model QFTs, $N(r)$ can be expressed in a simple way in terms of  the field operators $\phi(r)$ of a quantum field, $N(r) = \phi^*(r) \, \phi(r)$. Since Pearle and Squires \cite{PS94} have argued for choosing the collapse rate proportional to the mass, a better alternative to eq.\ \eqref{LambdaN} may be to replace the particle number density operators by the mass density operators.)

This model is as hard to distinguish experimentally from standard QFT as the GRW model from standard quantum mechanics. Thus, contrary to what is sometimes conveyed in the literature, it is not a matter of necessity to employ a CSL model for dealing with QFT. I would rather say that the generalization to QFT of the GRW model with flashes is simple and natural.

This model does not allow superluminal signalling by means of nonlocal correlations, and does not lead to any divergences. Ref.~\cite{ssr} includes a discussion of superselection rules in this model.

Where is the collapsing wave function in this model? With every time $t$ there is associated a wave function
\begin{equation}
  \psi_t =\frac{W_{t-t_n}\, K_{n} \, \psi}{\bigl\| W_{t-t_n}\, K_{n} \, \psi\bigr\|} \,,
\end{equation}
where $n$ is the number of flashes between $t_0$ and $t$, all of which enter into $K_{n}$. This function $\psi_t$ is what is best regarded as the wave function at time $t$; it is called the \emph{conditional wave function}, since it depends on the realization of the flashes between $t_0$ and $t$. It is also called the \emph{collapsed wave function} since each flash contributes a factor to $K_{n}$ that collapses $\psi$ (in one variable) to a neighborhood of the flash. As time proceeds, $\psi_t$ collapses whenever a flash occurs, and evolves deterministically in between. As a consequence of \eqref{ndistr}, the distribution of the flashes after $t$, conditional on the flashes between $t_0$ and $t$, is given by the formula \eqref{ndistr} for $\PPP$ with $\psi$ replaced by $\psi_t$ and $t_0$ replaced by $t$.

Another fact I would like to mention for later reference is that the probability that no flash occurs before time $t$ is given by 
\begin{equation}\label{noflash}
  \|W_{t-t_0} \, \psi\|^2\,.
\end{equation}

\section{Relativity}

A relativistic quantum theory without observers must overcome the difficulty of reconciling relativity with the quantum nonlocality discovered by Bell \cite{Bellbook}, which asserts that sometimes spacelike-separated events influence each other. 

\begin{figure}[ht]
\includegraphics[width=.4 \textwidth]{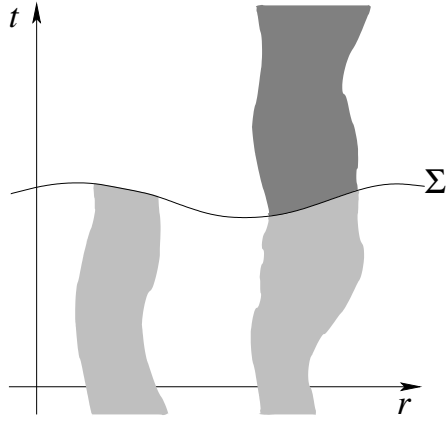}
\caption{Given a foliation of space-time into spacelike 3-surfaces, one can define a version of the GRW model with matter density ontology, in which a collapse of the wave function leads to a discontinuous change of the matter density $m(r,t)$ (shades of grey) along a 3-surface $\Sigma$ from the foliation.}
\label{figfour}
\end{figure}

A radical explanation of nonlocality postulates the existence of a preferred slicing, or \emph{foliation} (as mathematicians would say), of space-time into spacelike 3-surfaces. This foliation might naturally arise from the space-time geometry; for example, it might consist of the 3-surfaces of constant timelike distance from the big bang. Assuming a preferred foliation, a simple and natural extension of Bohmian mechanics to relativistic space-time has been worked out by D\"urr et al.~\cite{HBD}, and something analogous can presumably be done for the GRW model \cite{Mau05}, be it with the matter density or with the flash ontology, see Fig.~\ref{figfour}.

While the existence of a preferred foliation seems a possibility worth considering, the conventional understanding of relativity forbids it. It is hard to make precise what is allowed and what is not, but it would seem that special 3-surfaces playing the role of simultaneity-at-a-distance are not allowed. If a preferred foliation exists, the conventional understanding of relativity is wrong. This motivates the search for alternatives, i.e., for relativistic theories that do not require special foliations or any other additional structure in space-time providing a synchronization of spacelike-separated events. Such an alternative is not in sight with Bohmian mechanics, but such a version of GRW with flashes is possible! I will describe one in the next section. This confirms a hope expressed by Bell in 1987 \cite{Belljumps}:
\begin{quotation}
  I am particularly struck by the fact that the [GRW] model is as Lorentz
  invariant as it could be in the non-relativistic version. It takes away the 
  ground of my fear that any exact formulation of quantum mechanics
  must conflict with fundamental Lorentz invariance.
\end{quotation}

The observation that induced Bell to write this was the multi-time translation invariance of the original GRW model: Consider two quantum systems at a great distance from each other (far greater than their sizes); suppose the two systems are entangled but do not interact, $H= H_1 + H_2$, and that each is adequately described by nonrelativistic quantum mechanics. Then the effect of a small Lorentz boost is, to first order, a relative time translation; that is, the two systems get shifted in time by different amounts. In other words, exactly the synchronization (of events in one system with those in the other) that a preferred foliation would provide gets changed. The claim is that the joint distribution of the flashes is covariant under relative time translations, provided the two systems have different types of flashes. To see this, one may start from the following formula (here it suffices to consider $N=2$): It is the analog of formula \eqref{ndistr} for $N$ noninteracting systems with different types of flashes, and provides the joint distribution of the first $n_1, \ldots, n_N$ flashes with label $1,\ldots,N$. We denote by $X_{i,k}$ the $k$-th flash of type $i$.
\begin{equation}\label{multidist}
  \PPP\Bigl( X_{i,k} \in \D^4x_{i,k} : i\leq N, \: k\leq n_i \Bigr) =
  \Bigl\| \bigotimes_{i=1}^N K_{i,n_i} \,{\psi}\Bigr\|^2 \, 
  \prod_{i=1}^N \prod_{k=1}^{n_i} \D^4x_{i,k}\,,
\end{equation}
where the operator $K_{i,n}$ is given by \eqref{Kdef} and \eqref{Wdef} with $\Lambda$ replaced by $\Lambda_i$ and $H$ replaced by $H_i$. Now consider a pattern of flashes and apply a time translation to system 1 alone, shifting all flashes of type 1 to the past by some amount $\Delta > 0$ but leaving the flashes of type 2 unchanged. Apply this shift to the probability distribution and conditionalize on those $n'$ flashes that end up in the past of $t_0$; then this distribution is given again by \eqref{multidist} but with $\psi$ replaced by the conditional wave function
\begin{equation}
  \psi_\Delta = \frac{W_{\Delta+t_0-t_{n'}}\, K_{1,n'} \, \psi}
  {\bigl\| W_{\Delta+t_0-t_{n'}}\, K_{1,n'} \, \psi\bigr\|}\,.
\end{equation}

\section{A Relativistic GRW Model}

I now present a way of making the GRW model Lorentz invariant \cite{Tum06a}, exploiting the flash ontology and the analogy with the multi-time translation invariance. The relativistic model is nonlocal, excludes superluminal signalling, and converges to the GRW model in the non-relativistic limit $c \to \infty$. The version I shall present treats $N$ distinguishable quantum-mechanical Dirac (relativistic spin-$\tfrac{1}{2}$) particles without interaction in either flat or curved relativistic background space-time.

\begin{figure}[ht]
\includegraphics[width=.4 \textwidth]{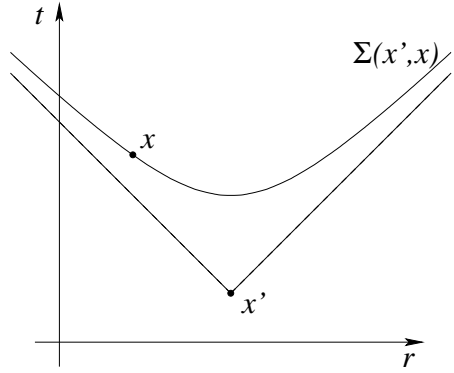}
\caption{The 3-surface $\Sigma(x',x)$ of constant timelike distance from $x'$ containing $x$}
\label{figtwo}
\end{figure}

Let me guide you through the equations defining the model. Each of the flashes wears a label $i\in \{1,\ldots,N\}$ corresponding to one of the $N$ variables of the wave function. 
The distribution of the flashes is again of the form \eqref{multidist}, but based on different $K$ operators, constructed in a Lorentz-invariant way. For two spacelike 3-surfaces $\Sigma_1,\Sigma_2$, let 
\begin{equation}
  U^{\Sigma_2}_{\Sigma_1}: L^2(\Sigma_1) \to L^2(\Sigma_2)
\end{equation}
denote the unitary evolution (according to the Dirac equation) from $\Sigma_1$ to $\Sigma_2$. For every spacelike 3-surface $\Sigma$, let $\Lambda_\Sigma(x)$ be the multiplication operator on $L^2(\Sigma)$ by a Gaussian centered at $x$,
\begin{equation}
  \Lambda_\Sigma(x) \, \psi(y) = \frac{\mathcal{N}}{\tau} \, 
  \exp \Bigl( - \frac{\text{s-dist}_\Sigma^2(x,y)}{2\sigma^2} \Bigr) \,
  \psi(y)
\end{equation}
where $y \in \Sigma$, ``s-dist$_\Sigma$'' denotes the spacelike distance along $\Sigma$, and $\sigma$ and $\tau$ are the same constants as before. The normalizing factor $\mathcal{N}$ is chosen so that 
\begin{equation}
  \int_\Sigma \D^3x\: \Lambda_\Sigma(x) = \frac{1}{\tau} 
\end{equation}
and may depend on $y$ if space-time is curved.
In the relativistic formula that will replace \eqref{Kdef}, we will use the expression $\Lambda_\Sigma(x)^{1/2}$ to replace the nonrelativistic operator $\Lambda(r)^{1/2}$, and we take $\Sigma$ to be the hyperboloid based at the previous flash, as in Fig.~\ref{figtwo}. This construction cannot be done with continuous collapse as in CSL models because it is the (previous) flash that defines the relevant 3-surface.

In detail, we write, for any sequence $f=(x_0,x_1,x_2,\ldots,x_n)$ of space-time points,
\begin{equation}
  K(f) = K_{x_{n-1}}(x_n) \cdots K_{x_1}(x_2) \, K_{x_0}(x_1)\,.
\end{equation}
The collapse operator $K_{x'}(x)$, which we define presently, corresponds to a flash at $x$ given that the previous flash was at $x'$, and is the analog of the nonrelativistic expression $K_{x'}(x) = \Lambda(r)^{1/2} \: W_{t-t'}$, except that we use the Heisenberg picture to define it on $L^2(\Sigma_0)$, with $\Sigma_0$ the 3-surface on which the initial wave function is specified (more about it at the beginning of the next section). We define
\begin{equation}\label{coldef}
  K_{x'}(x) = 1_{x\in \mathcal{F}(x')} \:
  \E^{-|x-x'|/2\tau} \: 
  U^{\Sigma_0}_{\Sigma} \:
  \Lambda_{\Sigma}(x)^{1/2} \: U^{\Sigma}_{\Sigma_0}\,,
\end{equation}
where $\mathcal{F}(x')$ means the future of $x'$ (the interior of the future light cone of $x'$), $|x-x'|$ the timelike distance between $x$ and $x'$, and  $\Sigma=\Sigma(x',x)=\{y \in \mathcal{F}(x'): |y-x'| = |x-x'|\}$ denotes the 3-surface of constant timelike distance from $x'$ containing $x$, as in Fig.~\ref{figtwo}. The first factor $1_{x \in \mathcal{F}(x')}$ ensures that flashes for the same particle are timelike separated. The second factor $\E^{-|x-x'|/2\tau}$ corresponds to the exponential distribution (with expectation $\tau$) of the proper waiting time $|x-x'|$. The last three terms are the Heisenberg-evolved version of $\Lambda_\Sigma(x)^{1/2}$. A key property of \eqref{coldef} is that
\begin{equation}\label{POVM}
  \int \D^4 x \: K_{x'}^*(x) \, K_{x'}(x) = 1\,.
\end{equation}

Since the distribution of the next flash depends on the previous flash, we need to specify a \emph{seed flash} $x_0$ as part of the initial condition; in fact, one seed flash $x_{i,0}$ for every particle label $i$. For the initial condition of the universe, they could be chosen to be, e.g., the Big Bang. With the notation $f_i=(x_{i,0},\ldots, x_{i,n_i})$ and
\begin{equation}\label{df}
  \D f = \prod_{i=1}^N \prod_{k=1}^{n_i} \D^4x_{i,k}\,, 
\end{equation}
the formula for the joint distribution of the first $n_i$ flashes of type $i$, the analog of \eqref{multidist}, reads
\begin{equation}\label{reldist}
  \PPP\Bigl( X_{i,k} \in \D^4x_{i,k} : i\leq N, \: k\leq n_i \Bigr) =
  \Bigl\| \bigotimes_{i=1}^N K(f_i)\,{\psi}\Bigr\|^2 \, 
  \D f\,.
\end{equation}
By \eqref{POVM}, this defines, at least formally, a consistent family of probability distributions.

This completes the definition of the model. Its Lorentz invariance is manifest from the fact that we did not mention a preferred slicing of space-time, or any other sort of additional space-time structure.

\section{Conditional Wave Function}

Do the seed flashes have to lie on the 3-surface $\Sigma_0$ on which we specify the initial wave function $\psi$? No. The measure $\PPP$ defined by \eqref{reldist} does not change if we replace $\Sigma_0$ by any other spacelike (Cauchy) 3-surface $\Sigma_0'$ and $\psi$ by $\psi' = U^{\Sigma_0'}_{\Sigma_0} \, \psi$, using the unitary evolution. Thus, as far as \eqref{reldist} is concerned, $\Sigma_0$ may lie in the past of some seed flashes and in the future of others, or even in the future of (some of) the random flashes governed by \eqref{reldist}. 

This should not be understood, however, as answering the question what function is best regarded as the collapsed wave function on a given 3-surface $\Sigma$. This is the question I address in this rather technical section. Before, however, I will provide a couple of facts that help us deal with the subtleties involved in conditioning on the information that a certain point $x$ was the last flash before a certain 3-surface $\Sigma$. First, the formula for the probability that no flashes occur up to the 3-surface $\Sigma$, i.e., the relativistic analog of \eqref{noflash}, reads
\begin{equation}\label{relnoflash}
  \Bigl\|\bigotimes_{i=1}^N W_{x_{i,0}}(\Sigma)\, \psi\Bigr\|^2
\end{equation}
with the abbreviation
\begin{equation}
  W_{x'}(\Sigma) = U_{\Sigma_0}^\Sigma\, \biggl(\int_{\mathcal{F}(\Sigma)} 
  \D^4x \: K_{x'}^*(x) \, K_{x'}(x) \biggr)^{1/2}\,,
\end{equation}
where $\mathcal{F}(\Sigma)= \cup_{x\in \Sigma} \mathcal{F}(x)$ is the future of $\Sigma$. And second, as a consequence of \eqref{relnoflash}, the probability of a particular sequence $f_i = (x_{i,0}, x_{i,1}, \ldots, x_{i,n_i})$ of flashes of type $i$ with $x_{i,1} \in \mathcal{F}(\Sigma)$, \emph{given} that the seed flash $x_{i,0}$ was the last before $\Sigma$, is not \eqref{reldist} but instead
\begin{equation}
  \frac{\bigl\| \bigotimes_{i} K(f_i)\,{\psi}\bigr\|^2}
  {\bigl\| \bigotimes_i W_{x_{i,0}}(\Sigma)\, \psi \bigr\|^2} \, 
  \D f\,. 
\end{equation}

Now I return to the question, what is the collapsing wave function (or conditional wave function) in this model? I have two alternative proposals. My first proposed wave function associated with any spacelike 3-surface $\Sigma$:
\begin{equation}\label{psiSigmadef}
  \psi_\Sigma =\frac{ \bigotimes_{i} 
  W_{x'_{i}}(\Sigma)\, K(f'_i) \, \psi}
  {\bigl\| \bigotimes_{i}  W_{x'_{i}}(\Sigma)\, K(f'_i) \, \psi\bigr\|}\,,
\end{equation}
where $f'_i$ is the finite sequence of all flashes of type $i$ up to $\Sigma$, and ${x}'_{i}$ is their last one. The distribution of the flashes  $\tilde{f}_i$ after $\Sigma$, conditional on the flashes $f'_i$ up to $\Sigma$, is given by 
\begin{equation}\label{distpsiSigma}
  \Bigl\| \bigotimes_i K(\tilde{f}_i) \, W_{x'_i}(\Sigma)^{-1} \, \psi_\Sigma \Bigr\|^2 
  \, \D \tilde{f}\,.
\end{equation}
That is, if we want to express the distribution of the future in terms of the present wave function, rather than of the initial wave function of the universe, then \eqref{distpsiSigma} is the more relevant formula than \eqref{reldist}.
As we push $\Sigma$ to the future, $\psi_\Sigma$ collapses whenever $\Sigma$ crosses a flash, and evolves deterministically in between; the evolution law reads
\begin{equation}\label{evolvepsiSigma}
  \psi_{\tilde\Sigma} = \frac{\bigotimes_i W_{\tilde{x}_{i}}(\tilde\Sigma)\, K(\tilde{f}_i) \, 
  W_{{x}'_i}(\Sigma)^{-1}\, \psi_\Sigma}{\bigl\| \bigotimes_i 
  W_{\tilde{x}_{i}}(\tilde\Sigma)\, K(\tilde{f}_i) \, 
  W_{{x}'_i}(\Sigma)^{-1}\, \psi_\Sigma \bigr\|} \,,
\end{equation}
if $\tilde\Sigma$ lies in the future of $\Sigma$, $\tilde{f}_i$ are the flashes between $\Sigma$ and $\tilde\Sigma$, and $\tilde{x}_i$ is their last one. It is perhaps worth emphasizing that, for any 3-surface $\Sigma$, to specify $\psi_\Sigma$ and the last flash before $\Sigma$ (of every type) suffices for determining the conditional distribution of all events in the future of $\Sigma$ given the past of $\Sigma$.

My second, inequivalent definition of a collapsing wave function is
\begin{equation}\label{phiSigmadef}
  \phi_\Sigma = 
  \frac{\bigotimes_i U_{\Sigma_0}^\Sigma\, K(f_i') \, \psi}
  {\bigl\| \bigotimes_i W_{x_i'}(\Sigma) \, K(f_i') \, \psi \bigr\|}
  = \bigotimes_i U_{\Sigma_0}^\Sigma\,
  W_{x_i'}(\Sigma)^{-1} \, \psi_\Sigma \,,
\end{equation}
which is not normalized, $\|\phi_\Sigma\|\neq 1$, and that is a feature one would not expect of a wave function; however, this definition leads to simpler formulas replacing \eqref{distpsiSigma} and \eqref{evolvepsiSigma}, namely
\begin{equation}
  \Bigl\| \bigotimes_i K(\tilde{f}_i) \, U^{\Sigma_0}_\Sigma\,
  \phi_\Sigma \Bigr\|^2 \, \D \tilde{f}
\end{equation}
and
\begin{equation}
  \phi_{\tilde{\Sigma}} = \frac{\bigotimes_i U_{\Sigma_0}^\Sigma\, 
  K(\tilde{f}_i) \, U^{\Sigma_0}_\Sigma\, \phi_\Sigma}
  {\bigl\| \bigotimes_i W_{\tilde{x}_i}(\tilde{\Sigma}) \, K(\tilde{f}_i) \, 
  U^{\Sigma_0}_\Sigma\, \phi_\Sigma \bigr\|} \,.
\end{equation}
The existence of inequivalent definitions of the wave function on $\Sigma$ underlines that the theory is not about wave functions, but that the wave function is a tool for formulating the law of flashes; see \cite{AGTZ} for further discussion of the status of the wave function.

If we associate with every $\Sigma$ a random wave function, be it by \eqref{psiSigmadef} or \eqref{phiSigmadef}, then to this ensemble of wave functions there corresponds a density matrix $\rho_\Sigma$. I have not succeeded so far in deriving a master equation for $\rho_\Sigma$, and the difficulties that arise suggest that none exists because the evolution of $\rho_\Sigma$ is not autonomous (i.e., there can be 3-surfaces $\Sigma_1, \Sigma_2$ and initial wave functions $\psi, \psi'$ such that $\rho_{\Sigma_1} = \rho'_{\Sigma_1}$ but $\rho_{\Sigma_2} \neq \rho'_{\Sigma_2}$). This may be part of the reason why previous work on relativistic collapse theories did not arrive at the present model: Di\'osi, for example, told me he expected that the first step towards a relativistic collapse model should be to find a relativistic master equation, but came to the conclusion that no such equation exists.

\section{Remarks}

An object moving at a speed $v$ close to the speed of light $c$ experiences fewer flashes, in fact by a factor of $1/\sqrt{1-v^2/c^2}$, as one would expect from time dilation.
No synchronization between the flashes for different particles is postulated or obtained in this model. 
With the matter density ontology, the model would not be Lorentz invariant. Instead, we would need a foliation (see Fig.~\ref{figfour}) to be able to write down a formula analogous to \eqref{mdef}. 

The reason why superluminal signalling is impossible is that, by repeated application of \eqref{POVM}, the marginal distribution of type-1 flashes,
\begin{equation}
  \int \D f_2 \, \bigl\|K(f_1) \otimes K(f_2) \,\psi\bigr\|^2 = 
  \bigl\| (K(f_1) \otimes 1) \, \psi \bigr\|^2 =
  \mathrm{tr} \bigl(\rho_1^\mathrm{red} \, K^*(f_1) \, K(f_1) \bigr)\,,
\end{equation}
does not depend on either $H_2$ or $\psi$ except through the reduced density matrix $\rho_1^\mathrm{red} = \mathrm{tr}_2 |\psi \rangle \langle \psi|$. Thus, for two separated systems, the distribution of the events in one region of space does not depend on the external fields, which we may imagine as arranged by an experimenter at will, in distant regions of space, and not either on features of the entangled wave function beyond those captured in the reduced density matrix.

The model is nonlocal. The basic reason is that distant flashes can be correlated, as 
\begin{equation}
  \bigl\|K(f_1) \otimes K(f_2)\,\psi\bigr\|^2 \neq \bigl\|K(f_1)\, \psi_1\bigr\|^2 \, 
  \bigl\|K(f_2) \,\psi_2\bigr\|^2 \quad \text{unless} \quad 
  \psi = \psi_1 \otimes \psi_2\,.
\end{equation}
While not all correlations are nonlocal correlations, there is nothing in the equations of the model that would preclude nonlocal correlations, for example between the space-time locations of two spacelike separated flashes. In this model we have, rather than merely the nonlocal probability law for results of experiments that we have in ordinary quantum mechanics, a nonlocal probability law for microscopic events, the flashes. An interesting feature of this model's way of reconciling nonlocality with relativity is that the superluminal influences between correlated events do not have a direction; in other words, it is not defined which of two events influenced the other. This is different in Bohmian mechanics with a preferred slicing, where, in the case of two experiments on entangled particles at spacelike separation, one is earlier than the other (in the sense defined by the slicing)---a fact that defines a direction of the nonlocal influence.

\section{Conclusions}

To conclude, the model describes a possible many-particle world in which outcomes of experiments have (to a sufficient degree of accuracy for all cases presently testable) the probabilities prescribed by quantum theory. It is fully compatible with relativity, in that it does not rely on a preferred slicing (foliation) of space-time. While the present model works on Minkowski space-time and Lorentzian manifolds, a relativistic collapse model on a \emph{lattice} ($\ZZZ^4$ instead of $\RRR^4$) was developed by Dowker and Henson \cite{dowker1}. Another model, which reproduces the quantum mechanical probabilities \emph{exactly}, is available in the framework of Bohmian mechanics \cite{HBD}, but uses a preferred slicing of space-time. Thus, with the presently available models we have the alternative: \emph{Either the conventional understanding of relativity is not right, or quantum mechanics is not exact.}

\bibliographystyle{aipproc}

\end{document}